\documentclass[letter,tradiabstract]{aa}

\usepackage{natbib}
\bibpunct{(}{)}{;}{a}{}{,}
\usepackage{graphicx}
\usepackage{txfonts}

\begin{document}

\title{Observational Consequences of the Recently Proposed Super-Earth Orbiting GJ\,436}

\author{J. L. Bean\inst{1} \and A. Seifahrt\inst{1}}

\institute{Institut f\"{u}r Astrophysik, Georg-August-Universit\"{a}t G\"{o}ttingen, Friedrich-Hund-Platz 1, 37077 G\"{o}ttingen, Germany\\
\email{bean@astro.physik.uni-goettingen.de, seifahrt@astro.physik.uni-goettingen.de}}

\date{Received 28 May 2008; accepted 19 June 2008}

\abstract{Ribas and collaborators have recently proposed that an additional, $\sim$\,5\,$M_{\oplus}$ planet orbits the transiting planet host star GJ\,436. Long-term dynamical interactions between the two planets leading to eccentricity excitation might provide an explanation for the transiting planet's unexpectedly large orbital eccentricity. In this paper we examine whether the existence of such a second planet is supported by the available observational data when the short-term interactions that would result from its presence are accounted for. We find that the model for the system suggested by Ribas and collaborators lead to predictions that are strongly inconsistent with the measured host star radial velocities, transiting planet primary and secondary eclipse times, and transiting planet orbital inclinations. A search for an alternative two planet model that is consistent with the data yields a number of plausible solutions, although no single one stands out as particularly unique by giving a significantly better fit to the data than the nominal single planet model. We conclude from this study that Ribas and collaborator's general hypothesis of an additional short-period planet in the GJ\,436 system is still plausible, but that there is not sufficient evidence to support their claim of a planet detection.}

\keywords{stars: individual: GJ\,436 -- planetary systems}

\maketitle

\section{Introduction}
The GJ\,436 system is unique among the nearly 250  extrasolar planetary systems identified so far\footnote{A regularly updated list of reported exoplanets can be found at http://exoplanet.eu.}. It contains a ``Hot Neptune'' planet (planet ``b'') that was originally discovered with high precision Doppler spectroscopy by \citet{butler04} and that was later found to transit by \citet{gillon07b}. This planet is the only known member of its class that transits its host star. Therefore, it is an interesting target for the particular investigations that are feasible for transiting exoplanets \citep[for a recent review of the observational techniques applicable to transiting exoplanets see][]{charbonneau07}. Since the discovery of the planet's transiting nature, follow-up studies have been carried out with the \textit{Spitzer Space Telescope} (hereafter referred to as \textit{Spitzer} for brevity) by \citet{gillon07a}, \citet{deming07}, \citet{demory07}, and \citet{southworth08}; and the \textit{Hubble Space Telescope} (\textit{HST}) by \citet{bean08}. 

The GJ\,436 system is also unusual because planet b has a significantly non-circular orbit despite its proximity to its host star. The planet has an orbital eccentricity $e$\,=\,0.15\,$\pm$\,0.01 as determined from analyzing the radial velocities of the host star with the constraint provided by the observed time of the planet's secondary eclipse that was observed with \textit{Spitzer} \citep{deming07,demory07}. An elliptical orbit for such a short-period planet ($P$\,=\,2.6\,d) is potentially at odds with the predictions of tidal theory, which suggests that the planet's orbit should become circularized on a timescale of $\lesssim$\,$10^{8}$\,yr \citep{maness07, deming07}.

\citet{maness07} found a significant linear trend in the radial velocities measured for GJ\,436 over 6 years superimposed on the signal from planet b. This discovery was interpreted to mean that GJ\,436 likely has an additional, but not necessarily planetary-mass, companion in a long-period orbit. \citet{maness07} investigated whether a long-period planet consistent with the radial velocity trend could be the perturber leading to excitation of planet b's orbital eccentricity. They found that it was possible, but far from certain owing to the unconstrained nature of the object causing the slow acceleration of GJ\,436. For example, \citet{maness07} suggested that a roughly Saturn mass planet in a 25\,yr orbit with $e$\,$\sim$\,0.2 would be consistent with the radial velocities and provide the necessary regular perturbations. 

Recently, \citet[][hereafter RFB]{ribas08} have proposed another explanation for GJ\,436b's high eccentricity. They suggest that an additional, short-period planet in the system would provide the necessary regular dynamical impulse to the transiting planet so that it would maintain its high orbital eccentricity in the face of tidal circularization over long timescales \citep[although this result has recently been questioned by][]{mardling08}. Such a planet also met their requirement to cause the transiting planet's orbital inclination to change by 0.1\degr\,yr$^{-1}$. They saw this change in inclination as the reason why \citet{butler04} didn't discover that GJ\,436b transited despite ostensibly having achieved the necessary photometric precision and sampling in their search. Their hypothesis is that the planet simply wasn't transiting at that epoch, while it was when observed at a later epoch by \citet{gillon07b} and the subsequent investigators mentioned above.

With the possible existence of another short-period planet in mind, RFB studied the radial velocities of GJ\,436 provided by \citet{maness07}. They identified a low-significance peak (20\% false alarm probability) in a periodogram of the single planet residuals and used that as the starting point for a two planet fit. Assuming Keplerian orbits, they were able to obtain a two planet model that fit the radial velocities significantly better than a single planet model.

The second planet in the RFB model has an orbital period, $P$\,=\,5.1859\,d, which puts it close to a 2:1 orbital resonance with the transiting planet, and minimum mass, $M\,sin\,i$\,=\,4.7\,$M_{\oplus}$. RFB proposed that this second planet exists based on the success of their model for providing a source for the transiting planet's eccentricity, a reason for the non-detection of transits by \citet{butler04}, and a fit to the radial velocities. If confirmed this would be the lowest mass planet yet found around a nearby, main sequence star. Therefore, the claim deserves further scrutiny.

One particular aspect of the RFB study that merits further investigation is the possible sensitivity of observables to gravitational interactions between the two planets in their model. These two planets would be in close, moderately eccentric, and possibly non-coplanar orbits and so their mutual perturbations might be significant on short timescales in addition to the long timescales that RFB only considered. If they are, then RFB's claimed detection might be spurious because their model did not account for them. The critical issue is that Keplerian orbits, which RFB used for their modeling of the radial velocities, are only strictly valid for the two-body problem (i.e. a single planet orbiting a single star and also in the absence of significant General Relativity effects). Such orbits are only a sufficient approximation for modeling the data for multi-planet systems when the planet-planet interaction timescales are much longer than the length of the observations. For systems where short-term interactions are occurring, or are even possible, a model based on direct integrations of the equations of motion (i.e. Newtonian orbits) should be calculated to check the validity of the Keplerian approximation. If the observables of interest would be significantly different when accounting for the interactions, then the Newtonian orbit model must be used.

In this paper we assess the consistency of RFB's model of the GJ\,436 planetary system with the observed host star radial velocities, transiting planet primary (transit) and secondary eclipse times, and transiting planet orbital inclinations when using Newtonian rather than Keplerian orbits. Pioneering work by \citet{laughlin01} and \citet{rivera01} have demonstrated that radial velocities with precisions on the order of a few m\,s$^{-1}$ are sensitive to short-term dynamical interactions for certain exoplanet systems. \citet{agol05} and \citet{holman05} have shown that transit timings measured with precisions of a few seconds up to a few minutes are quite sensitive to additional planets with masses down to even the terrestrial level. Transit timings for a planet near to a low-order resonance, as RFB propose for GJ\,436b, are particularly sensitive to very low-mass planets \citep{steffen05}.

\section{The model}
To generate a self-consistent planetary system model for comparing to observational data we used the Mercury code \citep{chambers99} to integrate the equations of motion. We assumed all the bodies were point masses and the only force considered was Newtonian gravity. We chose the Bulirsch-Stoer integration option as tests indicated this conservative method was necessary to achieve the desired accuracy.

The results of the integrations were recorded with a timestep of 0.01\,d and we used spline interpolation in the output grid of data to calculate model observables at arbitrary times. The combination of the selected sampling rate and interpolation method yielded accuracies of better than 0.02\,m\,s$^{-1}$ and 0.2\,s in the stellar radial velocities and planet transit times respectively for test cases. This level of accuracy is adequate because it is more than an order of magnitude better than the uncertainties in the observed data that was modeled.

\begin{table}
\begin{minipage}[t]{\columnwidth}
\renewcommand{\footnoterule}{}
\caption{Data from GJ\,436b eclipses}
\label{t1}
\begin{tabular}{lcc}
\hline
\hline
Parameter & Value & Source \\
\hline
Transit time (HJD) & 2454222.61564 $\pm$ 0.00060 & 1 \\
Transit time (HJD) & 2454280.78167 $\pm$ 0.00011 & 2 \\
Transit time (HJD) & 2454439.41607 $\pm$ 0.00068 & 3 \\
Transit time (HJD) & 2454444.70385 $\pm$ 0.00093 & 3 \\
Transit time (HJD) & 2454447.34757 $\pm$ 0.00080 & 3 \\
Transit time (HJD) & 2454463.20994 $\pm$ 0.00089 & 3 \\
Transit time (HJD) & 2454468.49911 $\pm$ 0.00094 & 3 \\
Secondary eclipse time (HJD)& 2454282.33 $\pm$ 0.01 & 4 \\
Inclination\footnote{At epoch 2454222.61564} ($\degr$) & 86.40 $\pm$ 0.10 & 1 \\
Inclination\footnote{At epoch 2454280.78167} ($\degr$) & 86.38 $\pm$ 0.10 & 2 \\
Inclination\footnote{At epoch 2454455.27924} ($\degr$) & 86.32 $\pm$ 0.08 & 3 \\
\hline
\end{tabular}
References: (1) Re-analysis of \citet{gillon07b} light curve; (2) re-analysis of \textit{Spitzer} data; (3) re-analysis of \textit{HST} data; (4) \citet{deming07}.\\[-15pt]
\end{minipage}
\end{table}

\section{The data}
\subsection{Radial velocities}
The radial velocities for GJ\,436 we analyzed come from \citet{maness07}. These same velocities were considered by RFB in their analysis using Keplerian orbits. We followed the suggestion of \citet{maness07} and added 1.9\,m\,s$^{-1}$ in quadrature with the reported errors to account for additional uncertainties arising from stellar and instrumental sources. Because GJ\,436 hosts a transiting planet its radial velocities will exhibit the Rossiter-McLaughlin (RM) effect during a transit \citep{rossiter24, mclaughlin24, gaudi07}. Therefore, these data cannot be used for orbit determination without including an additional model for the RM effect.

We searched the \citet{maness07} data for points that were possibly obtained during transit using the orbital ephemeris for planet b given by \citet{bean08}. We found that the observations with time stamps of 2453196.772 and 2453841.887\,HJD were obtained during a transit and we did not include these data in our analysis. The final data set of radial velocities we used contains 57 measurements spanning 6.5\,yr and having a typical uncertainty of 3.0\,m\,s$^{-1}$.

\subsection{Transiting planet parameters}
We also include in the dynamical analysis a combination of previously published and newly determined eclipse times and inclinations for GJ\,436b. We focused on the data that can be extracted from the transit light curves given by \citet{gillon07b} and those obtained with \textit{Spitzer} \citep{gillon07a,deming07} and \textit{HST} \citep{bean08}. Analyses based on different techniques and with different assumptions have been carried out on the transit light curves individually. We re-analyzed the photometry collectively with the same technique in order to obtain internally consistent data for the subsequent test of the RFB model\footnote{We chose to re-analyze the version of the \textit{Spitzer} reduced data presented by \citet{gillon07a} with the corrected time stamps (M. Gillon private communication, 2008).}. This step has the additional benefit of increasing the precision on the parameters of interest, which are time varying, by reducing their correlation with the physical parameters of the star and planet that the light curves are sensitive to, which do not vary with time.

We used the exact analytic formulae given by \citet{mandel02} to model the transit light curves and a Markov Chain Monte Carlo (MCMC) method similar to that of \citet{holman06} to identify the best fit model parameters. We assumed the planet and star radii were the same for each transit observation, but allowed each to have a unique central transit time and inclination. The \textit{HST} data include observations of five partial transits spread over 11 orbits of the planet and we determined a transit time for each one, but only one inclination for the group. There were 18 free parameters in total. We adopted the limb darkening constants for each light curve suggested by the authors in the corresponding papers initially presenting the data. We held fixed the transiting planet's orbital period to that determined by \citet[][2.643902\,d]{bean08}, and its orbital eccentricity, longitude of periastron, time of periastron, and velocity semiamplitude to those values determined by \citet{deming07}. We assume the mass of the star is 0.44 $\pm$ 0.04\,$M_{\sun}$, where the uncertainty in this value does not significantly contribute additional uncertainty to the transit times and inclinations. 

The identified parameter values and errors from the MCMC analysis were the median and 68\% confidence intervals respectively for aggregate of the trimmed chains. The transit times and inclinations used for testing the RFB model are given in Table~\ref{t1}. 

In addition to the re-determined transit parameters, we adopted the secondary eclipse time for planet b given by \citet{deming07} and \citet{demory07}. Those authors analyzed the same \textit{Spitzer} data but reached different conclusions regarding the uncertainty in the determined secondary eclipse time. We adopted the uncertainty given by \citet{deming07}, which is an order of magnitude larger than that given by \citet{demory07}. The secondary eclipse times reported by both groups are consistent within this larger range. The value we used in our analysis is given in Table~\ref{t1} for completeness.

\begin{table}
\caption{Single planet model parameters for GJ\,436}
\label{t2}
\begin{tabular}{lc}
\hline
\hline
Parameter & Value\\
\hline
$P$ (d) & 2.64390 $\pm$ 0.0000056  \\
$T_{c}$ (HJD) & 2454280.78168 $\pm$ 0.00011 \\
$T_{p}$ (HJD) & 2454280.221 $\pm$ 0.083 \\
$K$ (m\,s$^{-1}$) & 18.31 $\pm$ 0.57 \\
$e$ & 0.140 $\pm$ 0.007 \\
$\omega$ ($\degr$) & 357.6 $\pm$ 11.7 \\
$i$ ($\degr$)  &  86.36 $\pm$ 0.05 \\
$dv$/$dt$ (m\,s$^{-1}$\,yr$^{-1}$)  & 1.30 $\pm$ 0.27 \\
$\chi^{2}$  & 98.6 \\
Degrees of freedom & 60 \\
Radial velocity rms (m\,s$^{-1}$) & 3.8 \\
Transit time rms (d) & 0.00045 \\
Secondary eclipse time difference (d) & -0.008  \\
\hline
\end{tabular}
\end{table}

\section{Analysis}
\subsection{Evaluating the proposed Super-Earth planet}
Our evaluation of the RFB two planet model for the GJ\,436 system is based on the assumption that including an additional planet in a model for a system should yield a significantly better fit to the observational data for that system. To create a baseline for this test we first fit the data with a single planet model. We used a MCMC technique to determine the transiting planet's orbital parameters, a radial velocity linear trend, and a radial velocity offset to account for the relative nature of the radial velocities. There were eight free parameters in total. The parameters giving the best fit model and 1$\sigma$ confidence intervals from the resulting parameter distributions are given in Table~\ref{t2}. The $\chi^{2}$ of the best fit was 98.6 for 60 degrees of freedom. The larger than expected value for the corresponding reduced $\chi^{2}$ (1.6) is due to some small inconsistencies between the fit and the radial velocities, which is not unusual. The model predicted eclipse times and inclinations all fall within the uncertainties of the observed data.

With the fit-quality baseline established we then calculated the dynamical model radial velocities, eclipse times, and inclinations for the orbital parameters suggested by RFB and compared them to the observational data. We assumed that the parameters were osculating orbital parameters for the mean epoch of the radial velocities (HJD = 2453002.7). In addition to Keplerian orbital parameters, RFB solved for time varying components to the transiting planet's period, eccentricity and longitude of periastron. We did not fix these parameters in the model because if they are physical, then the dynamical calculations will naturally reproduce them. For a first test we just calculated the dynamical model with the nominal parameters suggested by RFB and determined the radial velocity offset that gave the best fit to the observations. We assumed that the inclination of the transiting planet was the average of the observed inclinations and the second planet was coplanar. The resulting $\chi^{2}$ was 4.4e6, which is five orders of magnitude larger than the reference one planet model. The rms of the radial velocity and transit time residuals was 11.5\,m\,s$^{-1}$ and 0.3\,d respectively. 

The orbital parameters given by RFB are not exact so we searched over the parameter space bounded by their given uncertainties while allowing the known planet's orbital parameters, second planet's eccentricity, velocity trend, and velocity data offset to vary. Using a combination of grid search and local minimization techniques we identified a best fit model that gave a $\chi^2$ of 144.9. This model had the second planet non-coplanar by 67$\degr$, which is unlikely to be a stable configuration. The smallest found $\chi^{2}$ for a coplanar model was 1.5e6. From this study we conclude that the specific two planet model proposed by RFB is completely inconsistent with the observational data.

\begin{figure}
\resizebox{\hsize}{!}{\includegraphics{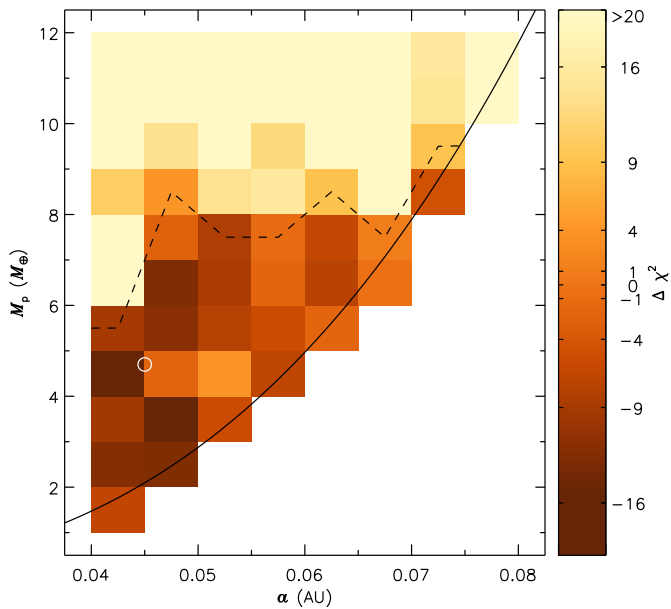}}
\caption{Intensity map of the $\chi^{2}$ surface for a two planet model fitted to the observed data. The shading indicates the difference in $\chi^{2}$ from the best fit single planet model. The dashed line gives the 3$\sigma$ upper limits for the mass of a second planet as a function of semimajor axis. The solid line delineates the lower mass limit for a second planet that RFB calculate can provide sufficient perturbations to excite the eccentricity of planet b. The circle shows the position of RFB's purported discovery.}
\label{f1}
\end{figure}

\subsection{Limits to additional planets}
While the currently available data can be used to rule out RFB's specifically proposed Super-Earth planet at very high confidence, what do they suggest regarding RFB's general hypothesis? To answer this question we examined how well the data could be fit by a two planet model where the mass and semimajor axis for the second planet was in the range RFB suggested was necessary to provide sufficient dynamical perturbations to planet b so that it maintains its orbital eccentricity in the face of tidal circularization over long timescales. We divided the region of interest in the mass -- semimajor axis plane of the second planet into 57 sub-regions. For each of these sub-regions we ran a local minimization algorithm initiated with some randomly selected orbital parameters for the second planet 100 different times and collected the results. For each run the orbital parameters for planet b were initiated at their single planet model best fit values given in Table~\ref{t2}. The mass and semimajor axis of the second planet were selected from, and restricted to, the sub-region limits. The eccentricity of the second planet was selected from the range 0.0 -- 0.3. The argument of periastron and mean anomaly of the second planet were selected from their fully defined ranges (both 0$\degr$ -- 360$\degr$). The second planet's inclination was selected out of, and restricted to, the range $\pm$15$\degr$ from planet b's inclination. The smallest $\chi^{2}$ found from the 100 different runs was taken to be representative of the best fit in the corresponding sub-region. We carried out 500 different runs in three of the sub-regions to verify that only 100 iterations were sufficient to reliably locate the best fit. The resulting fit-quality map of the second planet mass -- semimajor axis parameter space is displayed in Figure~\ref{f1}.

The results from this investigation indicate that there is a large range of possible parameters for a hypothetical second planet that give a better fit to the data than the single planet model. The identified best fit model has $\chi^{2}$ = 78.7 for 54 degrees of freedom. The second planet in this model has a mass $M_{c}$ = 5.0\,$M_{\oplus}$, and semimajor axis $a$ = 0.043\,AU. The false alarm probability (FAP) for the fit-quality improvement in this case is 5\%.

Examined in isolation the best fit solution could be considered as evidence to support RFB's specific claim for the discovery of a second planet in the GJ\,436 system because of its similarities with their model and in spite of the differences in the other orbital parameters. However, the consideration of planet-planet dynamical interactions has reduced the significance of the fit-quality improvement from that originally seen by RFB when using a model that does not include the interactions. The large FAP probability we found for the fit-quality improvement when incorporating a second planet in the model of system indicates that such an addition is not warranted. Furthermore, in the context of the larger parameter range investigation, we find that the best fit solution is not unique. There are solutions in other regions of the parameter space that are better than the single planet model and are statistically indistinguishable from the best fit. Therefore, we conclude that there is currently insufficient evidence to say that GJ\,436 definitely hosts a second planet in a close, exterior orbit to the already known planet, although the existence of such a planet cannot be ruled out.

Upper limits to an additional planet over the range of orbital semimajor axes considered can be set from the results of the fit-quality mapping by finding the point at which increasing the mass increases the $\chi^2$ above that of the single planet model more than a certain amount. The 3$\sigma$ confidence limit corresponds to an increase of 9 from the baseline model. This mass limit for a given semimajor axis is indicated in Figure~\ref{f1}. We find that the data can only rule out a second planet with semimajor axis similar to what RFB propose and $M_{c}$ $\ga$ 8\,$M_{\oplus}$ at high confidence.\\[-25pt]

\section{Summary}
We have shown that a Super-Earth planet like that one proposed by RFB could still exist in the GJ\,436 system, although their specific claim of a detection is erroneous. Recently, \citet{alonso08} carried out another investigation into the plausibility of RFB's proposed planet using a novel constraint on the allowable inclination change of the transiting planet and reached a
conclusion similar to ours. Ultimately, more observational data are needed to further constrain the architecture of the GJ\,436 system and its evolutionary history. RFB's proposed planet would have been the lowest-mass one yet discovered around a nearby star so the results of this work reemphasize the importance of considering planet-planet interactions when interpreting observations of multi-planet systems. Keplerian orbits are still a useful approximation for modeling many such systems, but their appropriateness for a particular case should always be tested.\\[-15pt]

\begin{acknowledgements}
J.B. and A.S. received funding for this work from the DFG through grant numbers GRK 1351 and RE 1664/4-1 respectively.\\[-25pt]
\end{acknowledgements}

\bibliographystyle{aa}
\bibliography{ms}

\end{document}